\documentclass[letterpaper]{article}
\usepackage{isea}
\usepackage{placeins}
\usepackage[pdftex]{graphicx}
\usepackage{times}
\usepackage{helvet}
\usepackage{courier}
\usepackage[numbers]{natbib}
\title{Situational Agency: The Framework for Designing Behavior in Agent-based art}
\author{Ary-Yue HUANG\textsuperscript{1}, Varvara Guljajeva\textsuperscript{23}\\
\textsuperscript{1} The Hong Kong University of Science and Technology (Guangzhou), China. yhuang948@connect.hkust-gz.edu.cn\\
 \textsuperscript{2} VCUarts Qatar. guljajevav@vcu.edu\\
\newline \textsuperscript{3} Academy of Media Arts Cologne, Germany.
\newline
}
\begin{document} 
\maketitle

\begin{abstract}
In the context of artificial life art and agent-based art, this paper draws on Simon Penny's {\itshape Aesthetic of Behavior} theory and Sofian Audry's discussions on behavior computation to examine how artists design agent behaviors and the ensuing aesthetic experiences. We advocate for integrating the environment in which agents operate as the context for behavioral design, positing that the environment emerges through continuous interactions among agents, audiences, and other entities, forming an evolving network of meanings generated by these interactions. Artists create contexts by deploying and guiding these computational systems, audience participation, and agent behaviors through artist strategies. This framework is developed by analysing two categories of agent-based artworks, exploring the intersection of computational systems, audience participation, and artistic strategies in creating aesthetic experiences. This paper seeks to provide a contextual foundation and framework for designing agents' behaviors by conducting a comparative study focused on behavioural design strategies by the artists.
\end{abstract}

Agent-based Art, Machine Learning, Behavior Design, Interactive, Artificial Life Art, Robotic Art, Media Art

\section{Introduction}
Artificial Life Art and Robotic Art are closely related fields that employ interactive agents with autonomous behavior. Agents can range from robots to virtual creatures, inhabit and explore within their environments, and interact with the audience or other living beings. To name a key early artworks incorporating agent behaviour, Gordon Pask's {\itshape Colloquy of Mobiles} (1968) demonstrated intersubjective interactions through communication between robots in what Pickering described as an ontological theatre\cite{Pickering}. William Latham's {\itshape Mutator}(1989) and Karl Sims' {\itshape Evolved Virtual Creatures}(1994) explored the evolution of agents' morphology and behaviors through evolutionary computation (EC)\cite{lai2021virtual,sims1994evolving}. Jon McCormack, Christa Sommerer and Laurent Mignonneau employed EC in the computational ecosystem (CE) to investigate the evolution of agents' behaviors and form within an ecological context \cite{McCormack2009,avolve}. Influenced by Rodney Brooks's Behavior-Based Robotics theory, which emphasizes direct coupling between perception and action, artists such as Ulrike Gabriel and Kenneth Rinaldo incorporated robotics in works like {\itshape Terrain}(1993-1997) and {\itshape Autopoiesis} (2000), allowing audiences to participate in and influence robotic behaviors\cite{metacreation}. Simon Penny's {\itshape Petit Mal} (1995) was designed as an autonomous robot that perceives and explores exhibition space while responding to audiences\cite{penny1998embodied}. Petra Gemeinboeck and Rob Saunders' Zwischenräume (2014) utilized machine learning (ML) to simulate curiosity, enabling robots to seek and develop aesthetic interests\cite{Petra2011}. We will utilize the term {\itshape agent-based art}, as used by Sofian Audry, to discuss artworks featuring agents with autonomous behavior, encompassing both virtual and physical media\cite{audry2021behavior}.

We aim to explore the practical knowledge of creating agent-based art to serve as a reference for artists and scholars interested in composing agent behaviour and this field in general. Next, we will introduce key concepts and examine historical discussions. This includes various perspectives on the aesthetics of agent behavior, frameworks for interacting with audiences, and the computational design of behaviors, all of which provide a foundation for discussing agent-based art.

\subsection{Aesthetics of Behavior and Autonomy}

Simon Penny uses the term {\itshape Aesthetics of Behavior} to describe this field as the interaction between culture and machine systems, emphasizing the creation of lifelike experiences through agents' behavior\cite{penny2000agents}. He explored this concept in {\itshape Petit Mal} (1995) artwork of his: an autonomous robot perceiving the audience's presence and responding accordingly. This embodied, user-manual-free interaction model stimulated the audience's curiosity and desire to engage. As audiences interacted with {\itshape Petit Mal} based on their associations with other living beings, Penny argues that these associations are inherently present within the cultural environment, viewing agents as cultural actors. Therefore, he considers the practical design of behavior and interaction, allowing audiences to project their metaphors and life experiences onto agents, as crucial factors in creating agents in artwork. 

Oliver Bown, Petra Gemeinboeck, and Rob Saunders analyze the aesthetics of interactive autonomous agents further. The continuously evolving characteristics of autonomous agent artworks challenge standard interaction models, such as prewritten scripts or mapped narratives\cite{BownAutonomous}. The examined artworks employ machine learning in their analysis, enabling agents to accumulate experience continuously. Agents may proactively seek interactions with the audience, develop independently, or engage in intimate dialogues. They argue that this evolving capability establishes autonomy within the audience's viewing and interactive experience, playing a crucial role in shaping interactions involving dialogue with the audience. Consequently, the audience is no longer merely controllers but active participants, co-creating a dynamic interactive relationship with the artwork.

\begin{figure*}[h!]
  \centering
  \includegraphics[width=\textwidth]{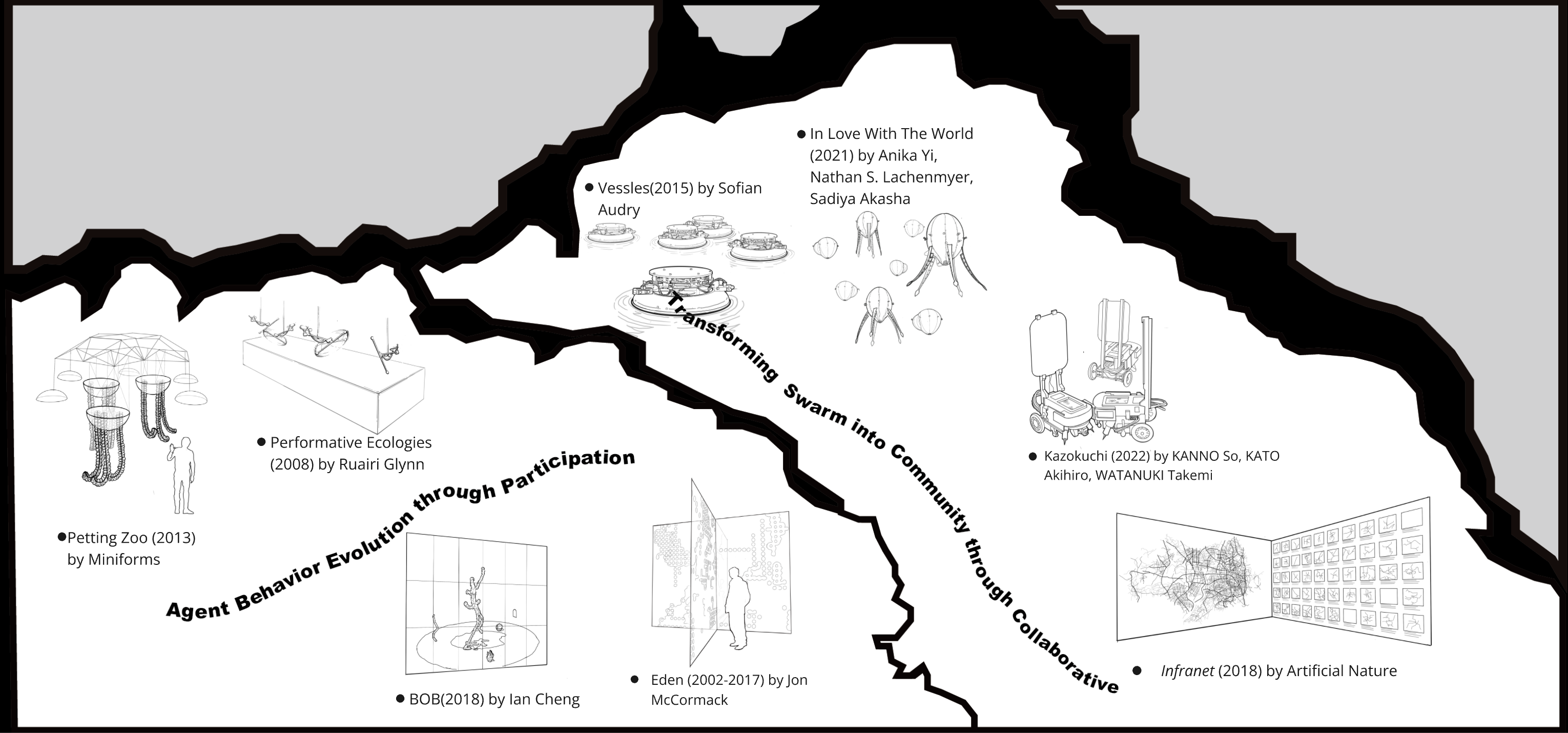}
  \caption{Two categories of agent-based artworks ©Yue HUANG}
  \label{fig1}
\end{figure*}

\subsection{Degrees of Behavior from Computation Perspective}
Sofian Audry analyzes the distinctions between various behavioral paradigms and their corresponding interactive aesthetics from the perspective of computational behavior design. He categorizes behavioral openness into 0\textsuperscript{th}, 1\textsuperscript{st}, and 2\textsuperscript{nd} order of behaviors, corresponding to stateless functions, rule-based behaviors, and adaptive or evolving behaviors, respectively \cite{audry2021artMLage}. These levels align with mapping, state machines, and machine learning techniques. Audry concentrates on 2\textsuperscript{nd} order of behavior, referring to the machine learning training process with a goal function as the aesthetics of adaptive behavior. His research provides a technical complement to Oliver Bown et al.'s arguments regarding agents' continuously evolving capabilities, also integrating the technical aspects of machine learning into the framework of Aesthetics of Behavior.
The concept of degrees of behavior serves as a foundation for analyzing both the computational systems involved in behavior design and the role of audience participation.  In the following chapters, we will discuss how evolving behavior systems with internal states respond to audience participation, which varies among artists' strategies in this research through analysing artworks.

\subsection{Motivation}
As practitioners of agent-based and interactive art, the motivations for this research arise from both domains. This study is driven by exploring how artists integrate emerging technologies into behavioral design, model agent behaviors to dynamically engage with audiences, and create compelling aesthetic experiences through agent-based art. In subsequent analysis, this paper proposes embedding the behavioral design of agents in their environments, conceptualizing the environment as an interactive semantic network composed of the audience, agents with technical systems, and other entities. Meaningful, situational interactions, and agent designs require the establishment of relationships, provision of context, and the stimulation of metaphors and associations. This conceptualization is encapsulated in the title of the paper, 'Situational Agency'. The study examines several significant works to develop this framework, analyzing the intersections between computational systems, audience participation, and artistic strategies to create behavioral aesthetics. Additionally, the paper compares various behavioral design and computational strategies and analyzes the interactive experiences of audiences with these artworks, as well as the diverse relationships formed between audiences and agents during interactions. These relationships further extend the paradigms of negotiation and communication proposed by Oliver Bown et al\cite{BownAutonomous}.

\section{Methodology}
Our research methodology combines archival research with practice-led investigation, situating selected artworks within an analytical framework for comparison and dialogue. These artworks are drawn from venues such as the SIGGRAPH archive, ISEA archive, Ars Electronica archive, VIDA (the VIDA Art and Artificial Life Awards), and ZKM. In this extensive map, we delineate two landscapes that are presented as two distinct categories: one where agents evolve their behavior through audience participation, and another where agents engage in collaboration and communication among themselves (Figure \ref{fig1}). Given the limited scope of this paper, we have excluded painting and musical performance robots, bio-robots, and bio-based neural network-driven robots, as their behavior design and aesthetic experiences differ.
In the following chapter discussion, we will elaborate on a dynamic framework for agents’ behavior design and refine it by incorporating theories of interactive art. The framework addresses the following key questions:

\begin{itemize}
\item In what ways do artists incorporate emerging technologies into the behavior design?
\item How can behavior be modeled to respond dynamically to audience participation?
\end{itemize}

\subsection{Agent and Environment}

The concept of agent aligns closely with the field of Embodied AI , a related research area that focuses on behavior learning and planning through interaction with environments \cite{lifeifeisurvey}. This alignment facilitates knowledge exchange and cross-disciplinary collaboration. We also use environment to describe the exhibition space where agents and audiences interact. In embodied AI, environments serve as training grounds for agents, typically involving the preparation of action datasets, the design of reward functions, or the use of imitation learning \cite{lifeifeisurvey}. In the interactive environments created by artists, while these techniques may be deployed, there is a greater emphasis on framing audience participation as training data for agent behavior learning or as feedback for the agent's reactions. Such interactions provide rich feedback for the development of agents. 

We propose to utilize ‘environment’ as a framework for evaluating agent behavior design because focusing solely on the behaviors of agents is insufficient.  Equally important is considering the broader environment in which these agents exist. Agents perceive, explore, interact with others, and develop behaviors based on feedback from interactions. This perspective aligns with Andrew Pickering's concept of the ontological theater, where he describes Gordon Pask's work as embodying a particular ontology of the world—not composed of static, independent entities, but rather woven from interconnected, dynamic systems that influence one another \cite{Pickering}.

Consequently, the artistic approach to behavior design must account for the environment in which the agent is situated. The artist's role can be envisioned as that of a situational creator, conceptualizing the agent’s environment as a network of dynamic relationships. The environment emerges through the continuous interactions among agents, audiences, and the evolving web of meanings generated by interaction. Artists design the forms perceived by the agent, its decision-making processes, and its responses to feedback. Furthermore, they decide how feedback becomes data for technical systems and guides audience engagement with the agent.

\section{Category 1: Evolution of agent behavior through participation}

\begin{figure*}[h]
  \centering
  \includegraphics[width=0.9\linewidth]{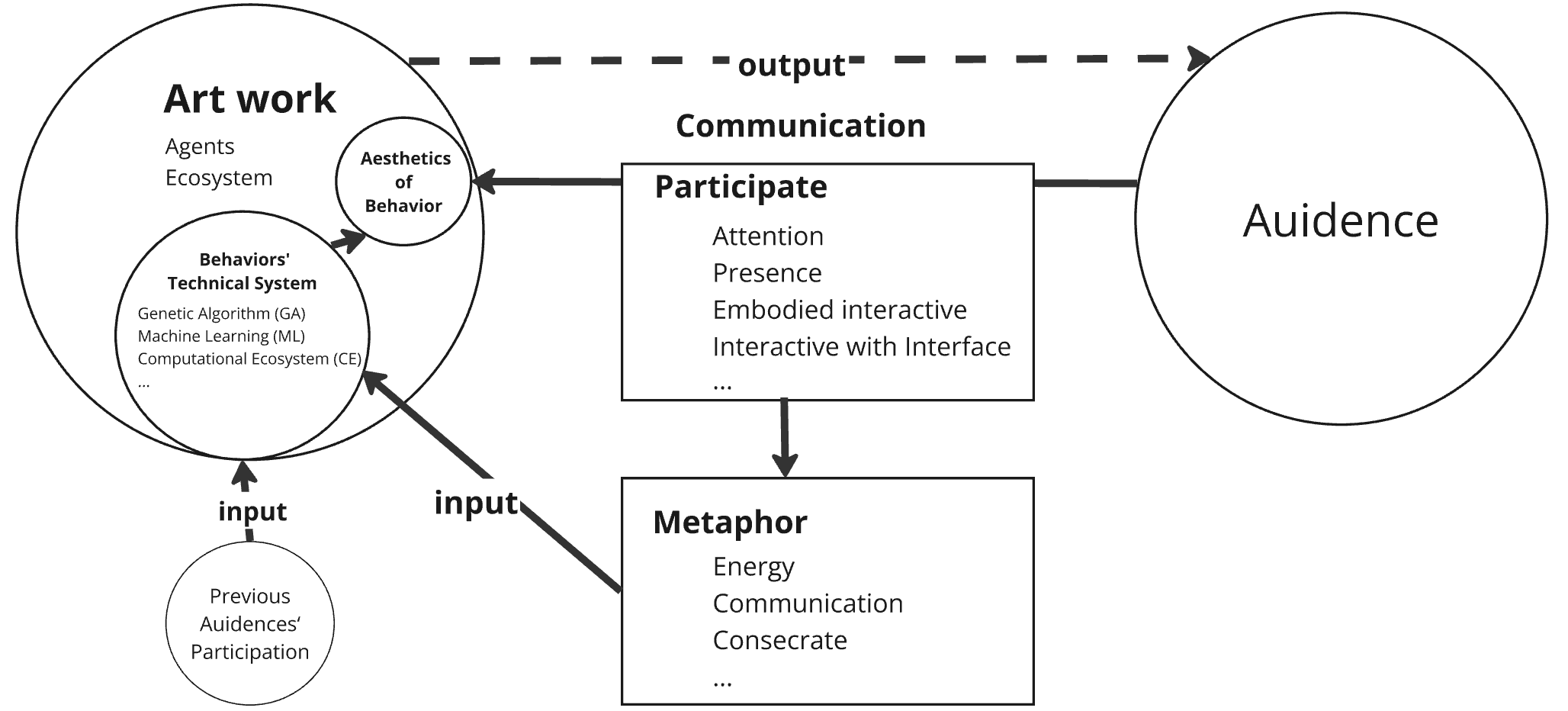}
  \caption{The relationships between audience participation, artworks, and artistic strategies in the category 1.©Yue HUANG}
  \label{fig2}
\end{figure*}

In this category, audience participation influences the evolution of artwork behavior. This interactive strategy aligns with the rhizomatic interaction strategy described by Ryszard W. Kluszczynski, in which audience interactions are not merely responded to once but persist within the system, generating multidimensional transformations \cite{kluszczynski2010strategies}. The way machine learning computes the transmission of audience participation closely corresponds to this aesthetic framework, as Sofian Audry suggests, with the agent's behavior accumulating internal experience over time \cite{audry2021artMLage}. The learning process inherent in machine learning also can be conceptualized as a form of processual aesthetics\cite{Yue}. However, further analysis is needed on how artists deploy technical systems and conceptualize audience engagement, as this is crucial for eliciting the metaphors and associations stored within the audience.

Strategies for artist-deployed audience participation in evolutionary behavior vary. The distinction lies in whether the agent's primary system is within a computational ecosystem or as an individual directly interacts with the audience in the exhibition space.  Computational ecosystems incorporate actors like inorganic matter, energy, and agents, creating a self-sufficient system. Integrating audience participation without disrupting stability is an artistic challenge. 

In Jon McCormack's {\itshape Eden}(2002-2017),  the agent emits sound to attract the audience's attention and harvests the energy released in computational ecosystem by the audience as both sustenance and energy for reproduction. Agents exhibiting favorable behavior will produce more offspring. When the audience loses interest, it triggers a behavioral mutation in the agent. McCormack drawing from his experiences in artistic contexts, predicts the audience's behavioral patterns and translates these patterns into meanings within computational ecosystems and Genetic Algorithms. Indirectly, the audience's behavior influences the agent's survival and reproduction rates. A similar strategy is employed in the work {\itshape Time of Doubles} (2011) by the duo of Artificial Nature and {\itshape A-volve} by Christa Sommerer and Laurent Minignnone, where the operation of CE functions independently of the audience's self-sufficiency \cite{Timeofdouble,avolve}. However, audience participation energizes the computational ecosystem, disrupting the energy field and invigorating the agents within it. Thus, in computational ecosystem, audience participation, through an intermediary, indirectly impacts the ecosystem and the evolutionary behavior of the agents.

Ruairi Glynn and Minimaforms' artwork conceptualizes audience participation as communication and negotiation between agents. Agents adjust their behavior by observing and interacting with audiences. Ruairi Glynn’s {\itshape Performative Ecologies} (2008) involves pole robots that attract audiences by performance and waving. These robots assess the level and direction of audience attention using OpenCV, score the performance, and evolve new actions through genetic algorithms \cite{glynn2008conversational}. {\itshape Petting Zoo} (2013) by Minimaforms employs spatial sensors allowing its tentacles to locate and interact with the audience, and record data to identify interaction patterns and adjust behavior\cite{Minimaforms}. When crowded, the tentacles display red to indicate agitation, but the audience can calm them to blue or cheer them to white by petting them. Both artworks evolve behavior through communication with another entity in real space.

Behavior modeling not only designs the actions but also the assessment and evaluation of input.  However, few works explore the evolution of these perception and decision systems. Ian Cheng's 2018 work {\itshape BOB (Bag of Beliefs)} is a notable example of such an exploration. BOB is a snake-like virtual creature capable of pruning, growing, and moving its body\cite{BOB}. Residing on the screen as a ‘forked snake,’ BOB continuously transforms itself through interactions with the audience. The audience can provide various odd items to BOB via an altar-like application and make comments such as ‘bombs are tasty’ or ‘fruits are delicious,’ resembling the kinds of lessons parents impart to their growing children. BOB may choose to trust a particular audience member and interact with the bomb they provide, but if the bomb destroys its body, BOB will lose trust in that individual and mark the bomb as a dangerous item in its internal state. Conversely, if a trusted audience member offers a delicious fruit, BOB will strengthen its belief that ‘fruit is tasty.’ These beliefs evolve as BOB continues to interact with its environment.

In this work, audience participation serves, to some extent, as data labeling, while BOB's neural network combines audience annotations with feedback from interactions with objects to inform its reasoning. Inspired by the research of Richard Evans\cite{evans2018learning}, BOB employs a reasoning engine capable of learning and deriving general rules from limited perceptual data. Given that, in real-world artistic contexts, audience interaction data may be insufficient for training neural networks that require large amounts of labeled data, Ian combined symbolic reasoning with machine learning, creating a hybrid network well-suited for the interactive scenarios he designed. In doing so, Ian outlined an artificial intelligence that possesses both motives and desires. BOB represents a figure that is both naive and sacred, with the audience alternating between the roles of believers and parents, cultivating a complex, trust-based relationship.

Each artwork establishes a unique context that creates meaning for participation and technical systems. For instance, Ruairi Glynn focuses on the behavior and learning of agents around performing with their bodies, where the exhibition space functions as a performance theater. Audience participation—specifically, their gaze—becomes contextualized as a form of performance evaluation, contributing to the agents' behavior-learning process. I have devised a diagram to illustrate context creation and the network within the environment (Figure \ref{fig2}). Artists conceptualize the audience's participation meaning (attention, presence, embodied interaction, etc.) and transmit it into life-related metaphors (energy, communication, consecrate, etc.). These are then input into evolutionary computing systems such as genetic algorithms or machine learning. These artistic strategies mobilize the audience's stored metaphors of interacting with life, establishing interactive context and situation. 

The interaction between the audience and the agent is not one of control, as in the reactive or mapping actions found in cybernetic art. Instead, the relationship is one of equal negotiation and exchange, where the agent evolves behavior through interaction with another living entity. Due to the inherent qualities of machine learning, which allow for the accumulation of experience and self-adjustment, each audience member's experience is never isolated. The interaction is not merely a reflection of the artist's design intentions; instead, each audience member's engagement contributes to shaping the agent's behavior. Every participant in the exhibition space plays a role in the agent's ongoing development.

\section{Category 2: Transforming Swarms into Communities through Collaboration}
\begin{figure*}[t]
  \centering
  \includegraphics[width=0.9\linewidth]{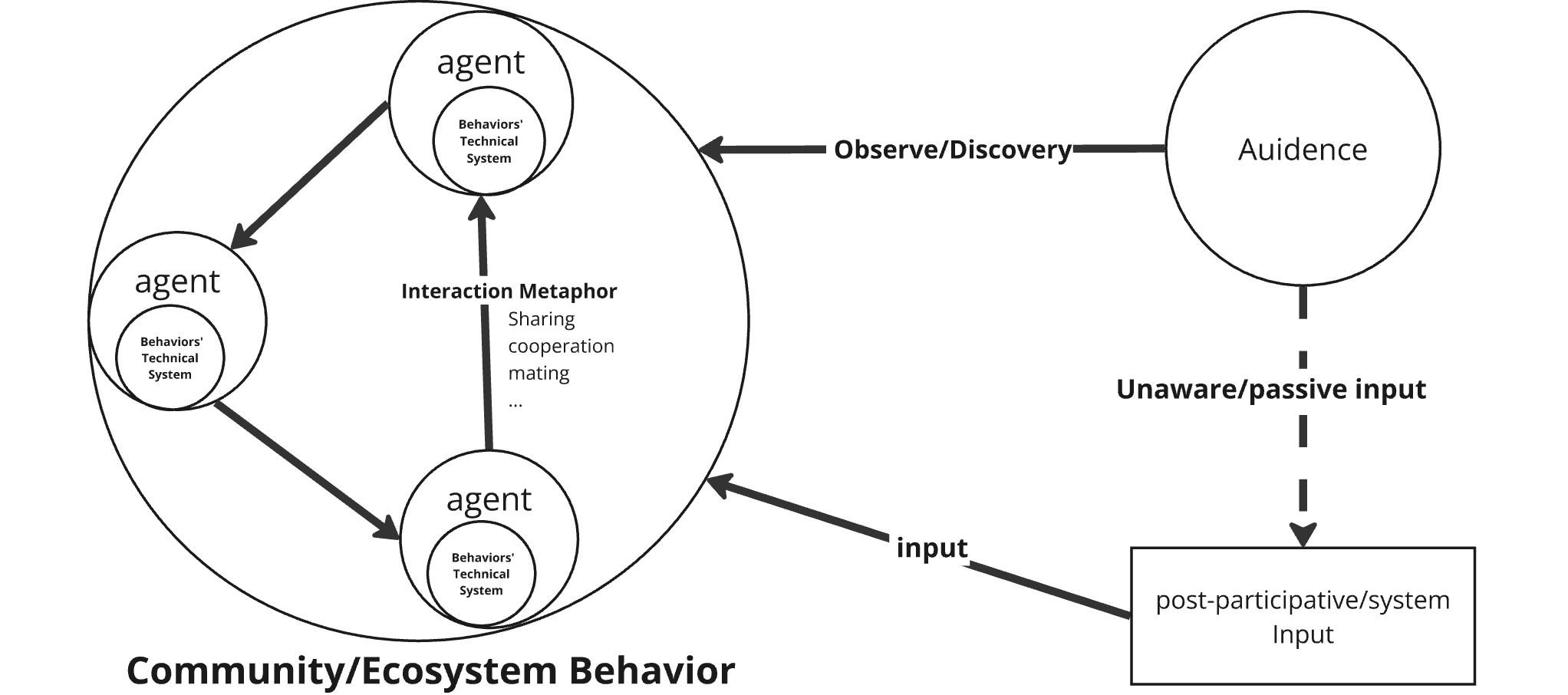}
  \caption{The relationships between audience participation, artwork, and artistic strategies in category 2.©Yue HUANG}
  \label{fig3}
\end{figure*}
Artworks in this category, though seeming to extend from Whitelaw's mention of bottom-up robotic art that group behavior emerges in basic individual behaviors\cite{metacreation}. These recent works integrate emergent technology to fabricate cooperation within the community, steering group behavior towards more organic and collaborative interactions. 

{\itshape Kazokuchi} (2022) by Kanno So, KATO Akihiro, and WATANUKI Takemi features a group of small robots (symbolizing houses), each carrying a screen displaying unique virtual creatures uploaded as tradable NFTs\cite{Kazokuchi}. Robots roam the space, interacting in the mating season as defined by artists.  Through physical interaction, virtual creatures between two houses may mate and upload children to the blockchain. This example exemplifies modeling swarm behavior by combining virtual creatures' behavior with the the physical body.

An artistic strategy to facilitate agent collaboration involves assigning distinct hobbies, flaws, and functions to different roles. Each agent is not omnipotent but requires collaboration with others possessing diverse capabilities to complete shared tasks or meet desires. {\itshape In Love With The World} (2021) by Anika Yi features the floating robots Xenojelly and Planula\cite{aquarium2022}. Despite its limited sensory range of 2.5 meters, Xenojelly endows the audience with additional capability and interest, such as collecting audience traces, which drives their behavior. When encountering peers, they share their collected traces and emerge from group behavior. Planula, having even weaker sensing, can collectively create a digital network that records traces from Xenojelly, serving as a basis for their actions and preserving and expanding their collective experiences.

Communication serves as a method for deploying cooperation among agents, as is observable in Anika Yi's works and Sofian Audry's {\itshape Vessels} (2015), where it serves as a data input for machine learning. Vessels is a group of 50 aquatic robots, each paired with different environmental sensors and sharing data nearby\cite{audry2021behavior}. Each robot utilizes reinforcement learning, allowing behaviors to evolve through continuously collecting environmental data and action feedback, further enhanced by peer-shared information. This facilitates meaningful group cooperation. Unlike individual organisms, social agents gain extra information through collaboration, which is then integrated into their operational mechanisms.

The sharing behavior between agents can also be interpreted as dynamic changes within their neural network structures. In {\itshape Infranet}(2018), Wakefield and Haru Ji conceptualize happiness as the internal state of agents, with sharing happiness viewed as habitual behavior. They deployed 4,000 to 8,000 agents, each nourished by urban map data. Neuro-Evolution of Augmenting Topologies (NEAT) provided each agent with a distinct neural network structure, determining its individual preferences for map data and driving the search for preferred routes\cite{Infranet}. Upon encountering a favored route, agents share their happiness with their peers, and receiving signals from agents with similar preferences increases their sense of happiness. Neighbors with lower happiness levels may accept these signals and, through mutation or network fusion, generate new neural structures that alter their tastes. In Infranet, agents exchange information not solely driven by survival but also by the desire to share happiness.

In these works, artists often prefer to invite audiences to observe the complex interactions and collective behaviors among agents, which are conceived as a community, rather than focusing on active participation. These ecosystem narratives depict agent communities, such as flying microbial entities that collect human signals and collective organisms living on city maps. Narratives define the agents' partners in learning and co-evolution, as well as the environments in which they reside. The focus of behavior design here shifts towards cooperation, communication, and the adaptation of behaviors between agents, with the audience assuming a more observational role (Figure \ref{fig3}). Communication between agents enters machine learning systems (e.g., NEAT, Genetic Algorithms, Reinforcement Learning) as feedback for learning, guided by the context to evoke mating, cooperation, and the sharing of happiness.

The audience's participation is more passive and implicit, often stemming from systems such as the trading of NFTs, city maps inhabited by humans, and walking paths traced by audiences. This aligns with Varvara Guljajeva's concept of {\itshape post-participation}, where human-generated data, often produced without the audience's awareness, becomes part of the technological system driving the artwork\cite{varpost-participation}. One of the aesthetics here emphasizes extending the interactive relationships to multiple interconnected systems. In agent-based art, {\itshape post-participation} is metaphorically applied to life-related aspects. City maps as sustenance or walking trajectories as references for agent movement. In this category, agents' group behaviors are sufficiently intricate that the audience may struggle to discern their underlying logic. This prompts them to observe and project their interpretations onto the agents' behaviors. The viewing experience transforms into something akin to observing wildlife, requiring careful analysis but rewarding in its discoveries. This form of participation evokes a relationship with species not closely related to humans, where we share certain aspects while maintaining a respectful distance.

\section{Discussion}
Previous discussions propose artistic strategies for constructing the aesthetics of behavior, where artists create situations by conceptualizing the environment where agents live as a dynamic network of relationships and meaning (see Figures \ref{fig2}-\ref{fig3}). We have examined how artists deploy technological systems and integrate audience participation into behavior design within the two categories discussed earlier, mapping out the relational networks. The construction of these situations prompts artists to reflect on the characteristics of the behavioral agents, guiding and shaping both the technological systems and audience engagement to present metaphors of life and direct the audience's projection of meaning, associations, and metaphors.

\begin{figure}[h!]
  \centering
  \includegraphics[width=\linewidth]{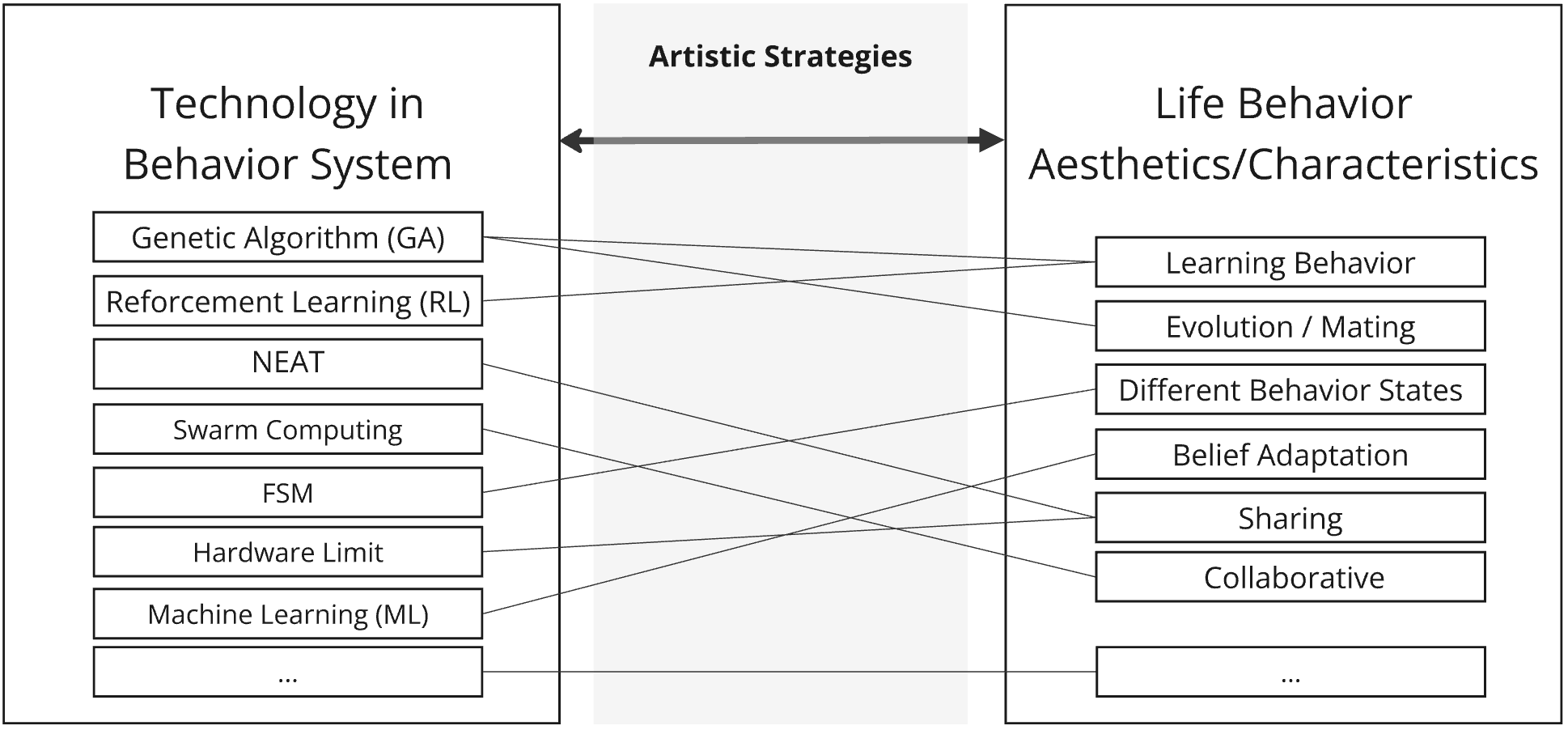}
  \caption{Artistic strategy guides technology systems to become aesthetics of behavior.©Yue HUANG}
  \label{fig4}
\end{figure}

One of the critical qualities of agent-based art is its emphasis on process, akin to Andrew Pickering's concept of the ontological theater. The process of interrelations unfolds through interactions and feedback loops between humans, robots, and non-human entities within the artwork. One key point is how computational technologies respond to audience participation. The deployment of technological systems is both technical and aesthetically significant. In the previous analysis, we saw how artists identify aesthetic qualities within technological systems and guide them toward the aesthetic of life-like behavior, linking these qualities to metaphors of life (Figure \ref{fig4}). For example, Finite State Machines (FSM) are often used to model different biological behaviors, and the learning process in machine learning, as Sofian Audry suggests, can be understood as a form of adaptive behavior. In artworks utilizing machine learning, the training process becomes part of the processual aesthetics, where agents accumulate experience through feedback from their interactions with the audience. The audience, in turn, may sense that the agent, as a system, is evolving its behavior through their participation, distinguishing this experience from non-systemic interactive art. 

\begin{figure*}[h!]
  \centering
  \includegraphics[width=\linewidth]{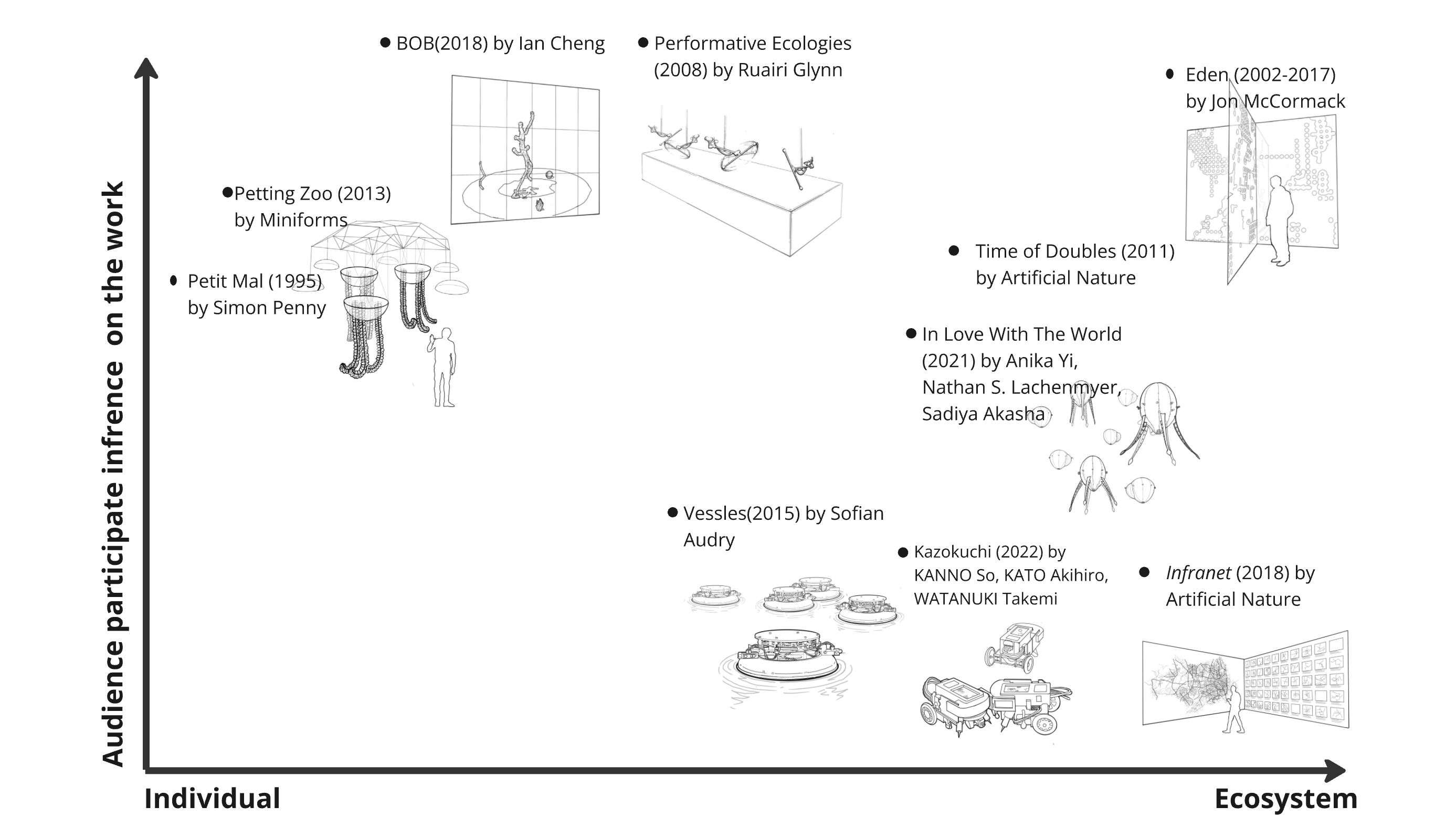}
  \caption{Mapping artworks by degree of audience participation: from individual impact to ecosystem influence.©Yue HUANG}
  \label{fig5}
\end{figure*}

However, the potential aesthetics of a technological system are not singular; the same technology can lead to different aesthetic expressions of behavior. For example, both {\itshape Eden} and {\itshape Performance Ecology} use genetic algorithms, but the former represents biological evolution, while the latter portrays behavioral learning through interaction with the audience. This difference arises from the distinct contexts and dimensions of life behaviors set by each work. Moreover, behavior can be influenced by implementing technological systems at various scales and aspects of life functions. In {\itshape BOB}, machine learning is used within the agent's belief system rather than behavior generation, causing actions to change as the agent's beliefs evolve. In {\itshape Infranet}, NEAT as machine learning model is deployed within the aesthetic taste system, allowing each agent's taste system to have dynamic neural networks of different structures, which can then share and mutate their structures.

Introducing audience participation into behavior modeling is an interactive art strategy and an indispensable aspect of agent behavior evolution within the environment. We subjectively map the analyzed artworks onto a chart, displaying the distribution of works along a spectrum from individual agents to ecosystems, with audience participation’s influence on the artwork’s evolution represented vertically. This chart illustrates the relationship and interactive experience between the audiences and the agents.
It can be observed, perhaps arbitrarily, that audience participation has a higher impact on the evolution of individual agents’ behavior. In this context, the audience can be seen as communicators, intimate partners, parents, or followers within the environment. Their interaction and negotiation with the agents facilitate behavioral changes. When considering the ecosystem case, audience participation is more indirect or implicit, visualized within the environment as energy, traces, and inanimate matter. It may result in ecosystems requiring balance, and agents focus more on interactions among themselves, forming group behaviors. The audience observes, projects, understands, and explores complex group behaviors. {\itshape Eden} (2002–2017) and {\itshape Time of Doubles} (2011) are exceptions, combining both forms of interactive experience, where the audience contributes to both the environment and the agents while also serving as observers of the ecosystem.

\section{Conclusion}

This paper extends Simon Penny's discussion on the aesthetics of behavior, exploring how artists design behaviors for agents in agent-based art. We propose that the design of the agent behaviors be considered within the context of the environment where these agents live, viewing the environment as a dynamic network. The artist's role is framed as one of situational construction. The article analyzes two specific categories of agent-based art to develop a dynamic analytical framework for examining how artists incorporate emerging technologies and audience participation into behavior modeling. Central to this framework is how artistic strategies guide and deploy these two key elements. Through a comparative study focused on behavior design strategies, this paper offers a context and framework for artists who wish to design behaviors for agents.

Furthermore, in an era when discussions surrounding artificial intelligence and robots are being reinvigorated, this research paints a landscape of how artists, driven by curiosity and practice, have constructed various, unfamiliar, and intriguing relationships between humans and AI/robots. These relationships, shaped by artistic guide interactive experiences and behavior design, allow humans to engage with these new entities in ways that go beyond treating them merely as tools.

%\section{Acknowledgments}
%The preparation of these instructions and the \LaTeX{} and Word files was facilitated by borrowing from similar documents used for ICCC proceedings.

\bibliographystyle{isea}
\bibliography{isea}

\section{Authors Biographies}
Ary-Yue Huang is an artist and researcher. His current work focuses on agent-based art and machine learning art—including software agents and physical robots—and he is interested in the performative aesthetics of agents. His previous artistic practice centered on speculative fiction, Film, and installation. He is currently pursuing a Ph.D. in Computational Media and Art at the Hong Kong University of Science and Technology (Guangzhou).

Prof Dr Varvara Guljajeva is an artist and researcher. She is an Associate Professor of Kinetic Imaging at VCUarts Qatar. Dr. Guljajeva has presented her research at major conferences like SIGGRAPH, CVPR, ISEA and more, with her work exhibited at Ars Electronica, ZKM, and Barbican.

\end{document}